\documentclass[conference]{IEEEtran}
\IEEEoverridecommandlockouts
\usepackage{cite}
\usepackage{amsmath,amssymb,amsfonts}
\usepackage{algorithmic}
\usepackage{graphicx}
\usepackage{textcomp}
\usepackage[table,xcdraw]{xcolor}
\usepackage{subfigure} 
 \usepackage{color}
 \usepackage{tabularray}
\usepackage{orcidlink}
\definecolor{MintGreen}{rgb}{0.603,1,0.6}
\definecolor{Silver}{rgb}{0.752,0.752,0.752}
\definecolor{HawkesBlue}{rgb}{0.796,0.807,0.984}
\definecolor{YourPink}{rgb}{1,0.8,0.788}

\def\BibTeX{{\rm B\kern-.05em{\sc i\kern-.025em b}\kern-.08em
    T\kern-.1667em\lower.7ex\hbox{E}\kern-.125emX}}

\begin{document}

\title{Design of a Teleoperated Robotic Bronchoscopy System for Peripheral Pulmonary Lesion Biopsy\\
\thanks{This work was supported by National Natural Science Foundation of China (U21A20480, 61950410618). Corresponding author: Lei Wang.}
\thanks{
$^{1}$ Research Centre for Medical Robotics and Minimally Invasive Surgical Devices, Shenzhen Institutes of Advanced Technology, Chinese Academy of Sciences, Shenzhen, 518055, China. 
	
$^{2}$ Shenzhen College of Advanced Technology, University of Chinese Academy of Sciences, Beijing 100049, China.

Emails:(xy.chen7; xh.xiong1; xm.wang2; p.li1; sm.wang, wk.duan; wj.du; tolu; omisore; wang.lei)@siat.ac.cn}
}

\author{Xing-Yu Chen$^{1}$$^{\orcidlink{0000-0003-1164-9537}}$,
Xiaohui Xiong$^{1,2}$$^{\orcidlink{0009-0003-1910-5101}}$, 
Xuemiao Wang$^{1}$$^{\orcidlink{0009-0006-3365-714X}}$, 
Peng Li$^{1}$$^{\orcidlink{0009-0007-4166-3029}}$, 
Shimei Wang$^{1,2}$$^{\orcidlink{0009-0004-9192-4539}}$,
\\Toluwanimi Akinyemi$^{1,2}$$^{\orcidlink{0000-0002-5598-8971}}$, 
Wenke Duan$^{1}$$^{\orcidlink{0000-0001-7509-7538}}$, 
Wenjing Du$^{1}$$^{\orcidlink{0000-0002-0571-3398}}$, 
Olatunji Omisore$^{1}$$^{\orcidlink{0000-0002-9740-5471}}$, 
and Lei Wang$^{1,*}$$^{\orcidlink{0000-0002-7033-9806}}$, }

\maketitle
\begin{abstract}
Bronchoscopy with transbronchial biopsy is a minimally invasive and effective method for early lung cancer intervention. Robot-assisted bronchoscopy offers improved precision, spatial flexibility, and reduced risk of cross-infection. This paper introduces a novel teleoperated robotic bronchoscopy system and a three-stage procedure designed for robot-assisted bronchoscopy. The robotic mechanism enables a clinical practice similar to traditional bronchoscopy, augmented by the control of a novel variable stiffness catheter for tissue sampling. A rapid prototype of the robotic system has been fully developed and validated through in-vivo experiments. The results demonstrate the potential of the proposed robotic bronchoscopy system and variable stiffness catheter in enhancing accuracy and safety during bronchoscopy procedures.

\end{abstract}


\section{Introduction}
\label{sec:introduction}

Lung cancer, ranking as the second most common cancer, poses a substantial global burden with high morbidity and mortality rates. Over the past five years, it has consistently accounted for an annual death toll of 1.9 million people~\cite{https://doi.org/10.3322/caac.21660}. This type of cancer usually start with the development of peripulmonary nodules in the lungs. Early detection and treatment of these pulmonary nodules are crucial for effective clinical intervention, leading to complete cures for many patients in the early stages and subsequently reducing lung cancer mortality rates~\cite{tanner2018standard}.

Conventional methods of lung biopsy typically utilize percutaneous or bronchoscopic needles~\cite{agrawal2020robotic}. Percutaneous needle biopsy involves inserting a needle into the lung lesion through the chest skin to obtain biopsy tissue. However, clinical practice and evaluation indicate that percutaneous needle biopsy carry a high risk of complications, including pneumothorax and significant bleeding~\cite{cicenia2020navigational}. As an alternative, transbronchial biopsy is gaining popularity due to its minimally invasive nature and relatively lower risks~\cite{wan2006bronchoscopic}.

Robotic technology, characterized by improved precision, spatial flexibility, and dexterity has the potential for enhancing minimal invasive surgeries. Robot-assisted minimal invasive surgeries (RAMIS) enable faster, safer, and more convenient navigation of surgical tools for intraluminal, endoluminal and transluminal interventions, eliminating the need for multiple or wide incisions~\cite{dupont2021decade}. The absence of large incisions during RAMIS provides numerous advantages, including small incision or no visible scarring, reduced postoperative pain, and the avoidance of general anesthesia for patients~\cite{zhou2013robotics}. Similarly, healthcare professionals can perform various interventions without been exposed to the operational risks~\cite{di2021medical}. 

In recent times, significant progress has been made in the field of robot-assisted bronchoscopy systems. A notable example is the Monarch\textsuperscript{\texttrademark} platform (Auris Health, Redwood City, USA). This system integrates an inner bronchoscope and an outer sheath equipped with electromagnetic (EM) technology for teleoperated navigation guidance. Another prominent bronchoscopy platform is the Ion\textsuperscript{\texttrademark} Endoluminal System (Intuitive Surgical, Sunnyvale, CA, USA) utilized for peripulmonary nodule interventions. Rather than using EM tracking, this system employs shape-sensing technology for tool navigation, using fiber Bragg grating and a video scope for navigation guidance during interventions. A common limitation of the Ion system is the absence of a direct visualization system during tissue sampling~\cite{kumar2021robotic}. Recently, the FDA-approved Galaxy\textsuperscript{\texttrademark} system, developed by Noah Medical (San Carlos, USA), has emerged, offering real-time navigation and lesion updates through tomosynthesis during lung interventions.

In addition to commercially available robotic bronchoscopy systems, noteworthy efforts have been made by academic researchers in the development of robots. Swaney et al. proposed a robotic system utilizing concentric tubes and a steerable needle with magnetic tracking, enabling precise movement through the bronchial wall~\cite{swaney2017toward}. Similarly, Amack et al. designed a concentric tubular robot with a compact, modular, and multi-stage mechanism to deploy a steerable needle through a standard flexible bronchoscope~\cite{amack2019design}. Duan et al. developed a bronchoscope robot with a small end-effector composed of a nickel-titanium tube, achieving three degrees of freedom in motion~\cite{duan2023novel}.

However, these efforts have predominantly focused on innovating mechanical structures with focus on novel designs for continuum robots or flexible robots. To address the problem of accurate biopsy in the highly dynamic peripheral pulmonary environment, this paper is aimed to present the design details of a robotic bronchoscopy system and characterization of its functionalities.  

The major contributions can be elaborated in the following two folds:
\begin{enumerate}
	\item variable stiffness catheter integrated in robotic bronchoscopy platform,
	\item a novel three-stage bronchoscopy procedure.
\end{enumerate}

The procedure follows initial insertion of a bronchoscope robot, dynamic adjustment of variable stiffness catheters, and robot assisted tissue sampling. All processes can be performed by remotely controlled robots. A rapid prototype of the robotic system has been fully developed and the feasibility and practicality are validated through in-vivo experiments.

\begin{table*}[h]
	\label{compare}
	\caption{Review and comparison with existing Robotic Bronchoscopy System.}
	\begin{tabular}{lcccccc}
		\hline
		System                                            & Control Method                                                                      & Navigation                                                                    & \begin{tabular}[c]{@{}c@{}}Bronchoscope\\ Size\end{tabular}                                                                 & \begin{tabular}[c]{@{}c@{}}Working Channel \\ Diameter\end{tabular} & \begin{tabular}[c]{@{}c@{}}Variable \\ Stiffness\end{tabular}       & \begin{tabular}[c]{@{}c@{}}FDA\\ Approval\end{tabular}               \\ \hline
		Monarch (Auris)~\cite{puchalski2021robotic}                            & Joystick-like controller                                                            & EM                                                               & \begin{tabular}[c]{@{}c@{}}Outer Sheath: 6.0 mm\\ Inner Scope: 4.2 mm\end{tabular}   & 2.1 mm                   &                           &\checkmark \\
		Ion (Intuitive)~\cite{puchalski2021robotic}                          & \begin{tabular}[c]{@{}c@{}}Console with scroll wheel \\ and track ball\end{tabular} & Shape Sensing                                                                 & 3.5 mm                                                                             & 2 mm                     &                           & \checkmark \\
		Galaxy (Noah)~\cite{Diddams2023RoboticBR}                            & Joystick-like controller                                                            & \begin{tabular}[c]{@{}c@{}}EM with C-arm \\ fluoroscopy \end{tabular} & 4.0 mm                                                                             & 2.1 mm                   &                           & \checkmark\\
		Duan et.al.~\cite{duan2023novel}~\cite{zhang2023design} & PC                                                                                  & EM                                                               & 3.3 mm                                                                             & 1.4 mm                   &\checkmark &                           \\
		Proposed System                                   & Multiple Tablets                                                                    & EM                                                         & \begin{tabular}[c]{@{}c@{}}Outer Scope: 5.2mm\\ Inner Catheter: 2.4mm\end{tabular} & 1.2 mm                   & \checkmark &                           \\ \hline
	\end{tabular}
	
\end{table*}

\section{Material and Methods}

The system comprises a teleoperated surgical robot specifically designed for trans-respiratory diagnosis, along with a corresponding master-slave control system. In contrast to existing robotic bronchoscopy systems, the proposed system integrates a commercial bronchoscope into the operating room to enhance cost efficiency and reduce usage complexity. The clinical procedure of the bronchoscopy remains unchanged, thereby shortening learning curve of the robot for surgeons.

\subsection{System Overview}
\label{systemdescription}

\begin{figure}[h]
	\centering{\includegraphics[width=0.8\columnwidth,height=0.6\columnwidth]{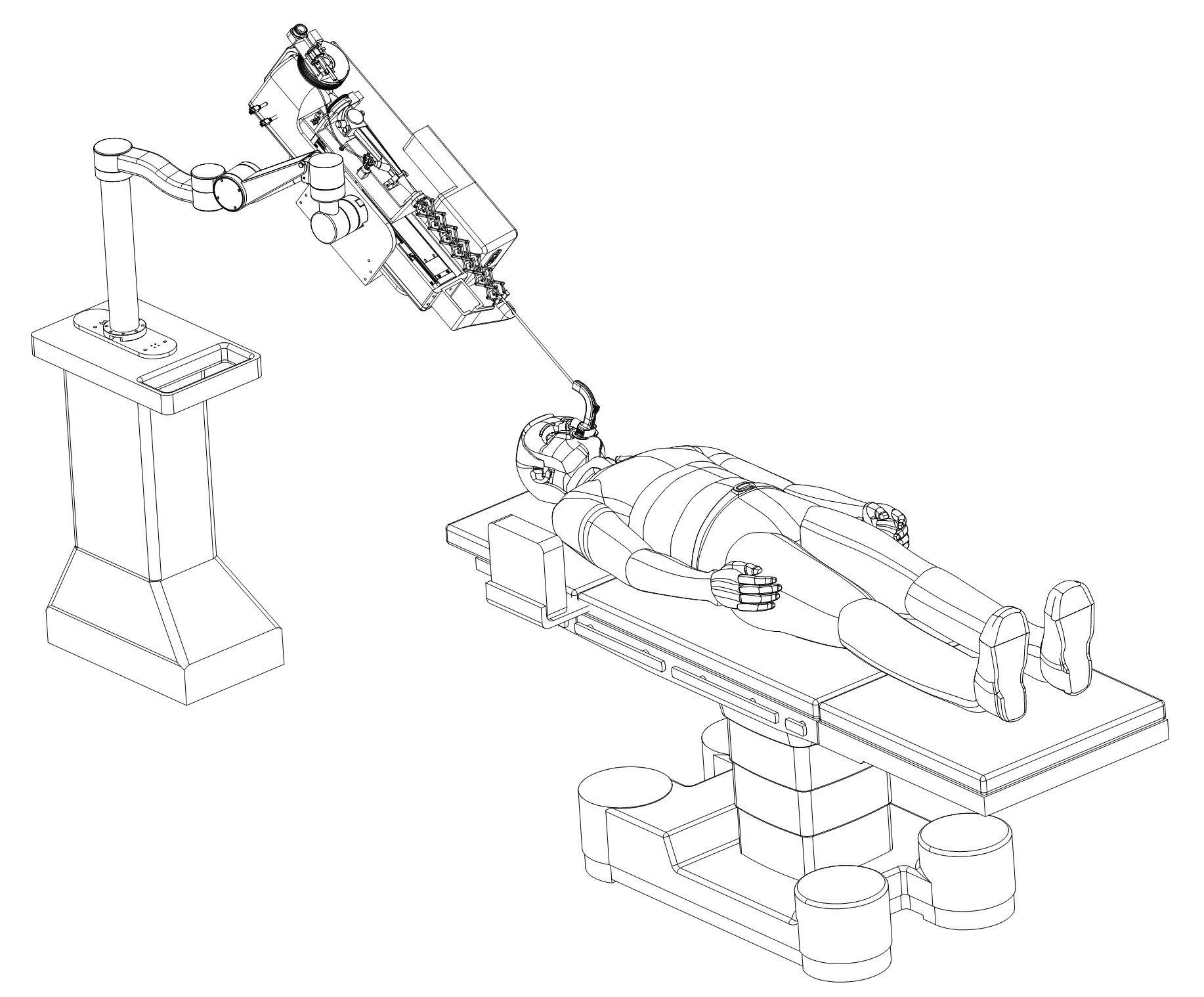}}
	\caption{Surgical scene of proposed bronchoscopy system mounted on a robotic arm, a trachea guide apparatus is inserted into patient's mouth to stabilize interventional direction.}
	\label{robot}
\end{figure}

Fig. \ref*{robot} illustrates the surgical scene of proposed robotic bronchoscopy system.
It involves mounting the bronchoscope manipulator on a passive robotic arm to ensure stability during surgery. In a typical bronchoscopy intervention, the manipulator is positioned adjacent to a patient, with the EM tracking system located nearby or under. The clinical procedure of the bronchoscopy remains unchanged, which allows the robotic system to simulate surgeons' realistic operating mode during the intervention.

\begin{figure*}[h]
	\centering{\includegraphics[width=1.2\columnwidth,height=0.8\columnwidth]{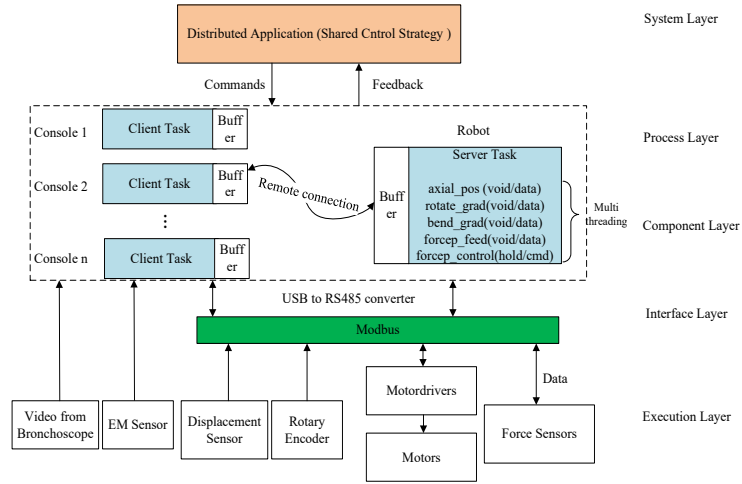}}
	\caption{Layers of the robotic bronchoscope system architecture.}
	\label{control_architecture}
\end{figure*}

The control architecture in Fig.\ref*{control_architecture} features master and slave devices with teleoperated control logic. Surgeons use tablets as master consoles (Microsoft Surface Pro 8) to teleoperate the bronchoscope robot, inserting it through the patient's trachea and utilizing the bronchoscope for advancements, rotations, and adjustment of the bending angle of the front end. Instructions from surgeons are wirelessly transmitted to the robot controller through the TCP/IP protocol. Subsequently, the robot responds to surgeons' commands by navigating the flexible bronchoscope along a specified trajectory. The robot also facilitates control of the VS-Catheters and biopsy forceps for advancement and tissue sampling through a 2.6 mm-diameter bronchoscope working channel. Employing a multi-operator strategy, the robot is controlled through a scheduling arrangement and weight distribution, allowing mentor surgeons and trainee surgeons to observe the same surgical site and collaboratively control the surgical instruments simultaneously.

\subsection{Bronchoscope Manipulator}

\begin{figure}[h]
	\centering{\includegraphics[width=0.8\columnwidth]{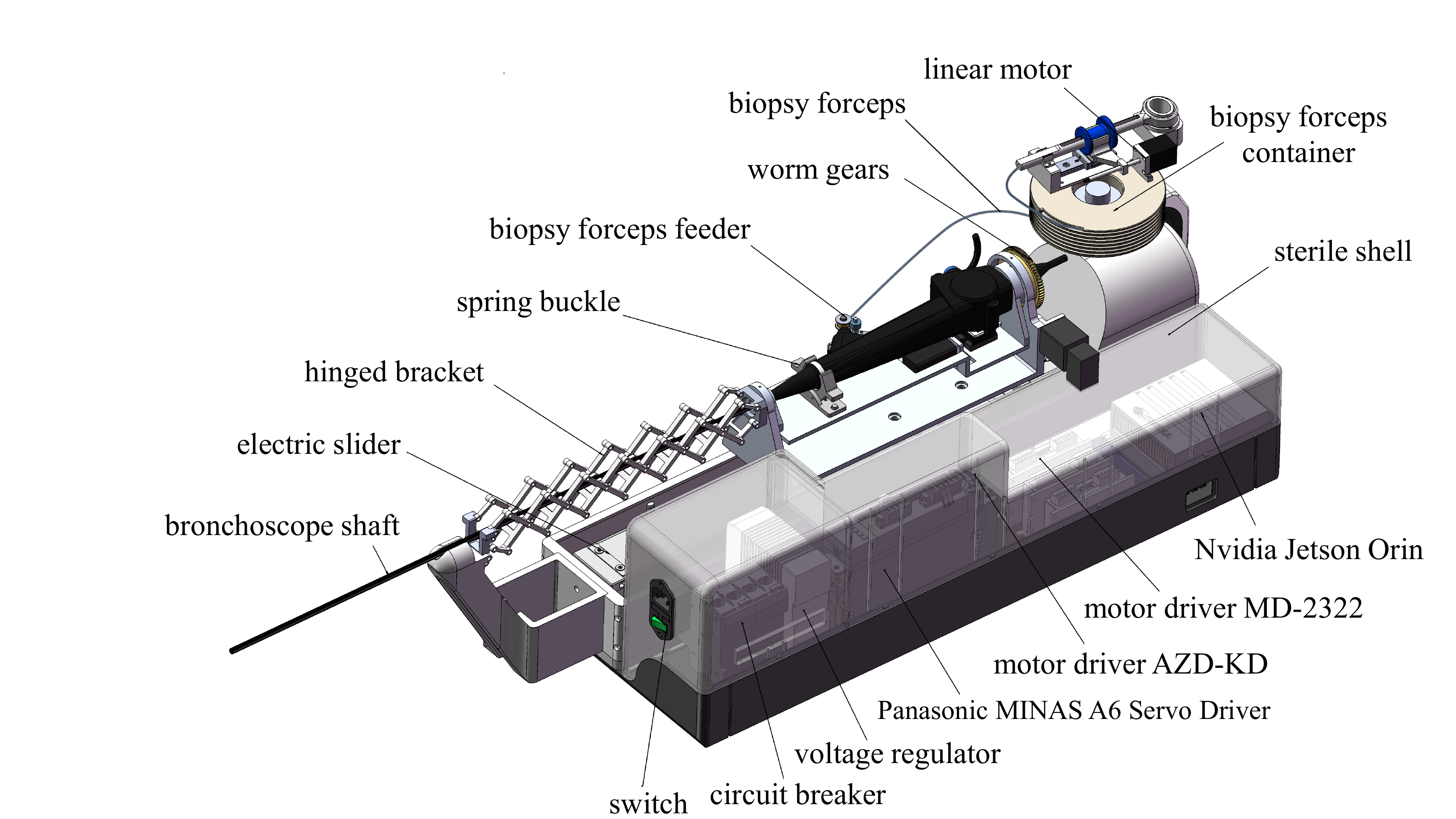}}
	\caption{Structure of the bronchoscope robot.}
	\label{structure}
\end{figure}

The design of the embedded system is based on our previous work in vascular interventional surgical robots~\cite{chen2022design}. Following the structure of bronchoscopy robot (in Fig. \ref*{structure}), we integrated a  Nvidia\textsuperscript{\textregistered} Jetson AGX Orin$^\text{TM}$ Developer Kit, which is programmed for controlling the electric slider (Panasonic\textsuperscript{\textregistered}, Osaka, Japan), rotary motor (Orientalmotor\textsuperscript{\textregistered}, Tokyo, Japan), motor drivers, gear sets, and biopsy forceps.
 
\begin{figure*}[!h]
	\centering{\includegraphics[width=1.2\columnwidth,height=0.6\columnwidth]{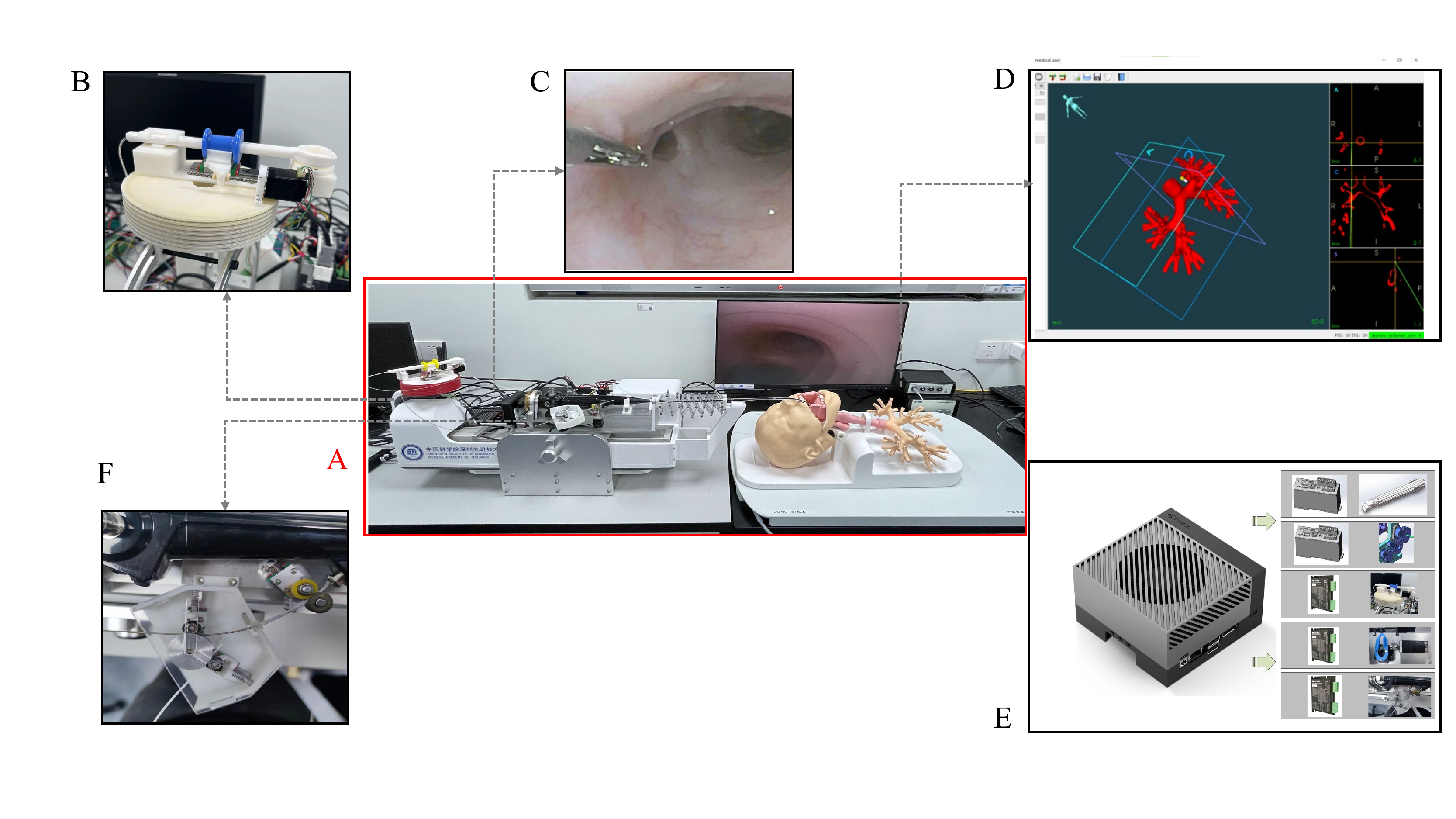}}
	\caption{Robotic Bronchoscopy System. (A) Robot and bronchus phantom; (B) Biopsy forceps manipulator; (C) Bronchoscopic video. (D) EM-Sensor based navigation system; (E) Embedded control system, electrical devices and motors; (F) Biopsy forceps and VS-Catheters delivery device.}
	\label{System}
\end{figure*}

The bronchoscope (UEWorld\textsuperscript{\texttrademark}, Xianju, China) utilized in our work can be readily replaced by other flexible endoscopes. This bronchoscope has an external diameter of 5.2 mm and a working channel of 2.6 mm, with the capability of a $160^{\circ}$ upward and $130^{\circ}$ downward bending angle. Commercial biopsy forceps can be installed in the forceps container. Several motors are mounted with a friction wheel to deliver biopsy forceps and VS-Catheters, as shown in Fig. \ref*{System}.

We have integrated 6-DoF EM sensors to enable real-time navigation, tracking, and three-dimensional localization of the surgical instruments. For this purpose, an EM tracking system (NDI Aurora, Waterloo, Canada) with a field generator that produces an EM field with known geometry is integrated with the robotic system. This provides information about the 6 degree of freedom position and orientation (roll, pitch, and yaw) of the EM sensor. To integrate endoscopic video and EM tracking information, a multimodal navigation system is employed, utilizing the open-source software CustusX~\cite{askeland2016custusx}. The software includes a toolbox of navigation features, image-processing algorithms, and connections to external hardware for image-guided therapy. By combining images, tracked surgical instruments, and computer display, a comprehensive navigation system is created, enabling real-time identification of the direction and position of the tip of the bronchoscope.

\subsection{Variable Stiffness Catheter}

\begin{figure}[h]
	\centering{\includegraphics[width=0.8\columnwidth,height=0.5\columnwidth]{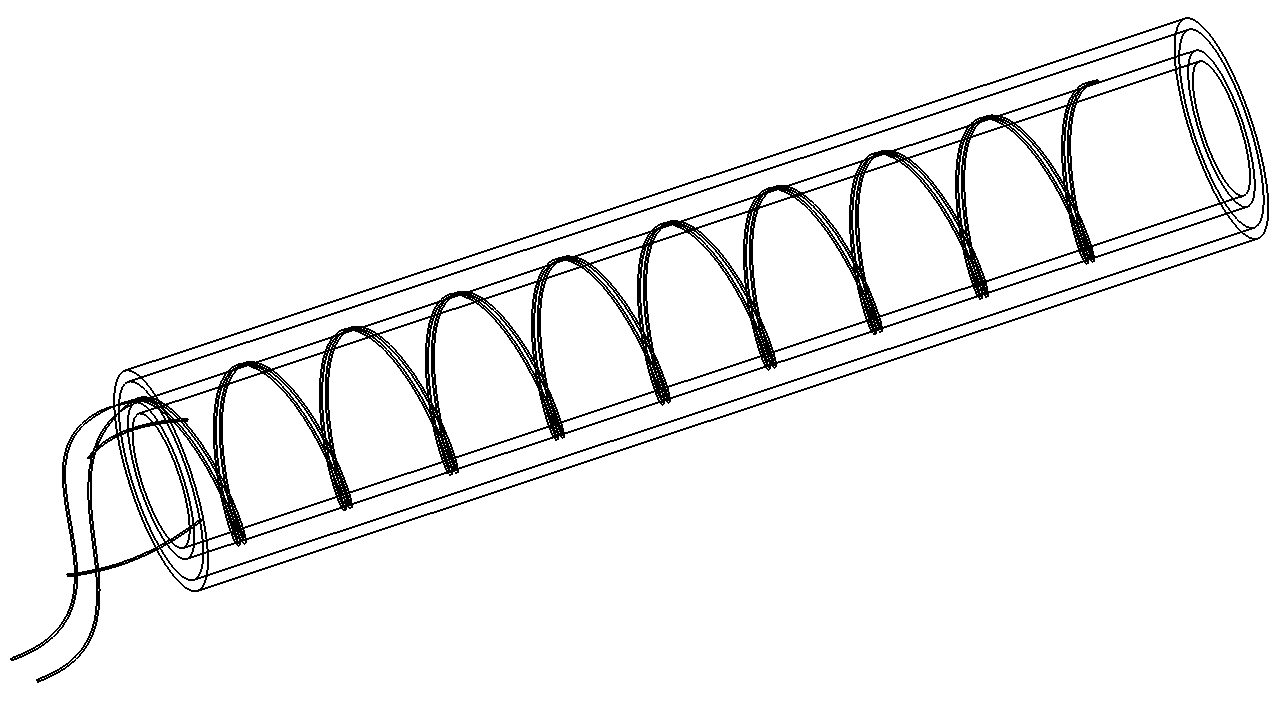}}
	\caption{Structure of proposed VS-Catheter, consist of hollow flexible tube fulfilled with LMPA, heating module, resistance measurement module and wire driving module.}
	\label{VSC}
\end{figure}

Catheters with the ability of variable stiffness are designed and utilized to facilitate the extension of surgical instruments through the working channel of bronchoscope, thereby reducing the potential risk of unintentional tissue penetration~\cite{chautems2020magnetic}. The catheter design aims to maintain optimal flexibility for seamless traversal of tortuous intraluminal pathways in the human body while also ensuring sufficient rigidity to provide adequate support during tissue biopsy. Specifically, we developed a novel variable stiffness catheter(VS-Catheter) shown in Fig. \ref*{VSC} composed of low melting point alloy (LMPA) to enhance the catheter's flexibility in the context of RAMIS and expand the accessible area for bronchoscopic biopsy forceps. 

Unlike the bronchoscope, the VS-Catheter can be inserted into narrower bronchi and offers more flexible control with dynamic stiffness. The VS-Catheter consists of a hollow flexible inner tube and an outer tube incorporating an interlayer infused with LMPA (Field's metal) at melting temperature of 47°C. The inner wall material of the VS-Catheter is made of medical-grade polytetrafluoroethylene (PTFE). This material has characteristics of low friction, heat resistance and chemical resistance, and is widely used in various interventional medical materials. The outer wall of the sheath is made of polyether block amide (PEBAX). Additionally, a heat-generating resistance wire is helically wound around the flexible inner tube. This wire serves the purpose of melting the Field's metal, thereby providing variable stiffness functionality to the catheter. The resistance wire and the Field's metal together constitute a stiffness measurement circuit, and a controller-based methodology is employed to achieve continuous adjustability of the catheter's stiffness. The wire embedded in the VS-Catheter is driven by a motor and can be pulled to control bending.

The catheter utilizes Field's metal phase change technology to achieve controllable stiffness. The cooling process of the catheter occurs naturally at the physiological temperature of the human body, eliminating the need for external stimuli and ensuring safety and reliability. The catheter incorporates two circuits: the heating circuit by current generator and the measurement circuit, enabling real-time adjustment of the catheter's stiffness to accommodate the intricate anatomical environment.

For safety reasons, alarm and protection methods are integrated into the control algorithm. Any abnormal resistance measurement triggers a warning to the main controller and interrupts the heating program. Two mechanisms are employed to secure the heating procedure: a fuse is installed in the heating circuit to prevent momentary high currents, and the intrinsic electrical characteristics of the VS-Catheter  prevent the accumulation of heat. Due to the higher impedance at the connection point between the VS-Catheter and the heating module, the heat generated in the tail area is also higher. When the accumulated heat becomes excessive, the heating circuit in the tail end of VS-Catheter will melt first, leading to the interruption of the entire heating circuit. Consequently, the catheter ceases to generate heat, avoiding excess heat accumulation and minimizing the risk to the patient.

\section{Novel Robotic Bronchoscopy Procedure}
\label{procedure}

\begin{figure*}[h]
	\centering{\includegraphics[width=1.15\columnwidth,height=0.64\columnwidth]{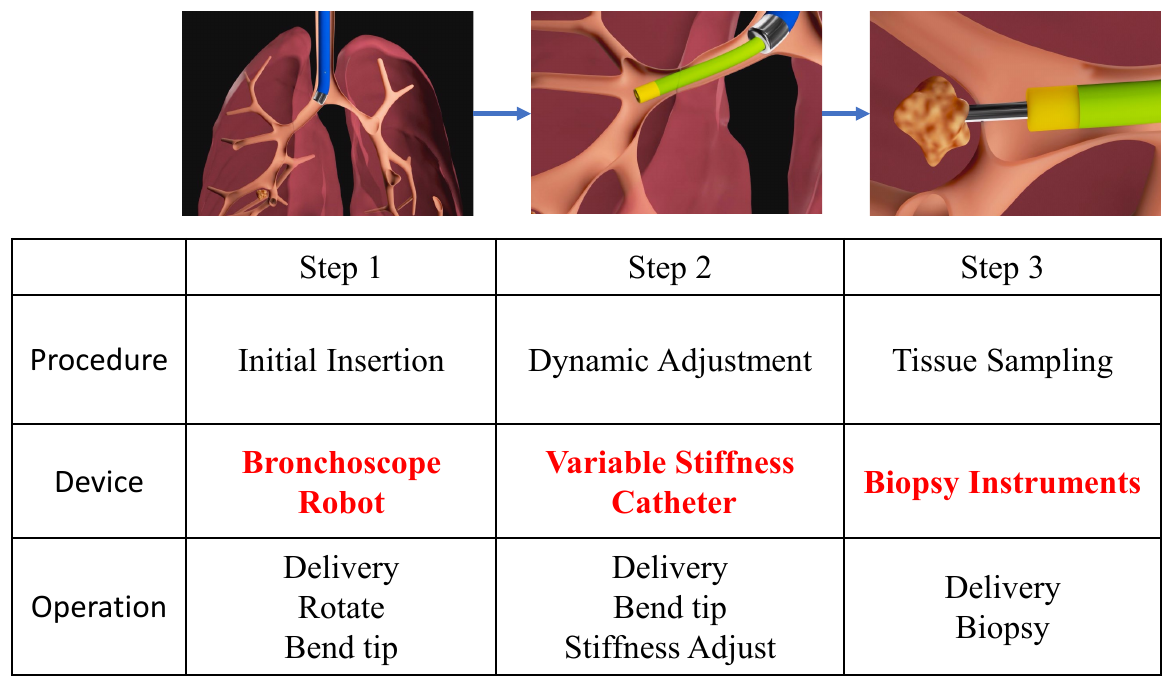}}
	\caption{Proposed three-stage bronchoscopy surgical procedures, with initial insertion, dynamic adjustment, and tissue sampling, composed of robotic bronchoscope, VS-Catheter, and tissue sampling by biopsy forceps, respectively.}
	\label{three_tools}
\end{figure*}

A novel three-stage robotic bronchoscopy procedure  proposed in Fig. \ref*{three_tools} involves the bronchoscope robot, VS-Catheter, and biopsy instruments. Table \ref*{table} presents essential mechanical parameters of the bronchoscopic tools. In the first stage, surgeons use tablets to navigate the bronchoscope robot to a target position, involving motions such as advancement, rotation, and tip bending.

In stage two, the VS-Catheter is heated (maintained at a safe temperature of 53°C) to decrease its stiffness. Subsequently, the softened VS-Catheter is inserted through the bronchoscope's working channel. Its dynamic stiffness can be adjusted based on factors such as heartbeat and respiration. Compared to the bronchoscope, the VS-Catheter can be inserted into narrower bronchi and offers more flexible control with driving wire. Once the softened catheter reaches the precise location of the targeted lesion, the current generator to the heating resistance wire is disconnected. Heat discontinuation allows the LMPA to cool naturally, restoring the catheter's stiffness.

Tissue sampling is performed using the biopsy forceps in the third stage. The surgeons can thereby examine the patient's airways, search for abnormalities or suspicious areas, and perform tissue biopsy for further testing. Surgeons have the capability to control all these three-stage processes using tablets and teleoperate the procedures. This emphasizes the importance of robotic surgery, as it enables surgeons to manipulate multiple medical tools without the need for shift changes, ensuring stability within the trachea.

\begin{table}
	\centering
	\caption{Essential Mechanical Parameters.}
	\scalebox{0.8}{
	\begin{tblr}{
			cell{2}{2} = {Silver},
			cell{3}{2} = {Silver},
			cell{4}{2} = {Silver},
			cell{5}{2} = {Silver},
			cell{6}{2} = {Silver},
			hlines,
			vlines,
		}
		Idx. & Tool          & Description                       & Value~ \\
		1    & Bronchoscope  & Outside Diameter                  & 5.2 mm \\
		2    & Bronchoscope  & Inside Diameter (Working Channel) & 2.6 mm \\
		3    & VS Catheter   & Outside Diameter                  & 2.4 mm    \\
		4    & VS Catheter   & Inside Diameter~                  & 1.2 mm    \\
		5   & Biopsy forcep & Diameter~                         & 1.0 mm    
	\end{tblr}
}
\label{table}
\end{table}

\section{Evaluation and Experiment}

\subsection{Evaluation of System}

The delivery accuracy of proposed robotic bronchoscopy system is quantitatively measured, including the delivery of only the VS-Catheter, the delivery of only the biopsy forceps, the rapid and slow delivery of the biopsy forceps within VS-Catheter. Each category underwent 20 sets of experiments. The encoder data of the motor connected to the friction wheel, i.e., the number of motor steps, were recorded by motor-driver. When the motor steps change one, the motor rotates 0.36°, allowing calculation of the motor rotation angle. The friction wheel diameter is 18mm, so the delivery distance for rotation of friction wheel can be calculated.
\begin{figure}[h]
	\centering{\includegraphics[width=0.9\columnwidth,height=0.6\columnwidth]{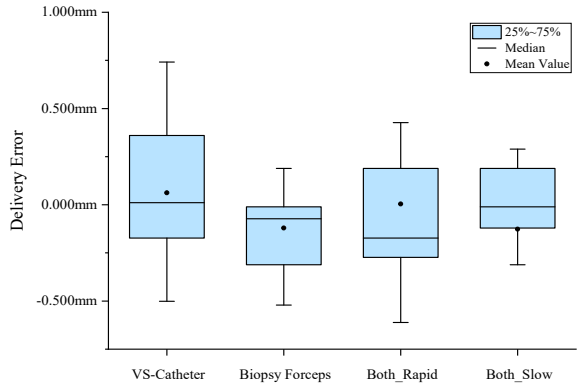}}
	\caption{Delivery accuracy of VS-Catheter and biopsy forceps.}
	\label{accuracy}
\end{figure}
The maximum delivery error for the VS-Catheter was 0.741mm, while for the biopsy forceps, it was 0.520mm. This discrepancy arises from the rigidity of the biopsy forceps compared to the flexibility of the VS-Catheter, leading to some deformation during delivery. Moreover, when delivering the biopsy forceps inside the VS-Catheter, the error noticeably increases compared to delivering without the VS-Catheter. The solution to this issue is to reduce the delivery speed to one-fifth of the previous speed (delivery speed from 36°/s reduced to 7.2°/s), the delivery accuracy remains under 0.5mm.
%

\begin{figure}[h]
	\centering{\includegraphics[width=0.9\columnwidth,height=0.6\columnwidth]{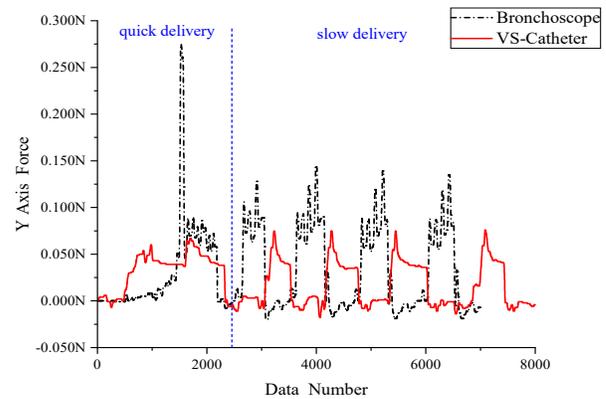}}
	\caption{Force measurement of bronchoscope and VS-Catheter.}
\end{figure}

Force experiments are also conducted in the bronchus phantom. A three-dimensional force sensor was installed beneath the bronchus phantom to measure forces in the X, Y and Z directions. The forces caused by the bronchoscope and the VS-Catheter within the phantom are measured. The bronchoscope and the VS-Catheter are each delivered five times within the phantom, including one fast delivery and four slow deliveries. The slow speed is 1/5 of the fast speed. The force caused by the bronchoscope is significantly larger than that of the VS-Catheter, reaching 0.3N during fast delivery. However, the force exerted by the VS-Catheter within the phantom remained relatively stable at 0.075N, regardless of the delivery speed. This also demonstrates that using the VS-Catheter inside the bronchus, as opposed to using the bronchoscope, results in less impact on the patient's bronchus.

\subsection{In-vivo Experiment}
%

In-vivo animal experiment is carried out on a swine with a weight of 30 kg. The studies are conducted to assess the robot's design concept and feasibility for transbronchial biopsy. The complete procedures of robot-assisted bronchoscopy involve several steps, including acquisition of imaging data, segmentation, registration, and 3D construction of the bronchus. The process begins with acquiring imaging data through CT scan, to create a 3D map of the swine's airways. This map is integrated into the system to provide a visual representation of the surgical workspace. After the registration is complete, the surgeon utilizes the 3D map to plan bronchoscopy procedure.

\begin{figure}[h]
	\centering{\includegraphics[width=1.0\columnwidth,height=0.4\columnwidth]{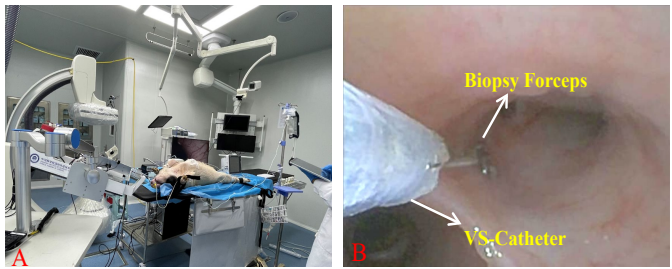}}
	\caption{Proposed bronchoscopy system. (A) In-vivo experiment scene; (B) Intra-body navigation view of the bronchoscopy tools in swine's trachea.}
	\label{animal}
\end{figure}

After setting up the robotic manipulator on the passive arm and calibrating it, the surgeon remotely controls the robotic system with the multimodal navigation system. The surgeon identifies location of suspicious areas of tumors or lesions(we pick a random location as target), and determines the optimal route in bronchus to reach them.  As shown in Fig. \ref*{animal}, the robotic system is positioned next to the animal. The system is controlled by two surgeons using tablets. Fiducial EM sensor is attached to the swine's forebreast, and another tracking EM sensor on the tip of the bronchoscope. In the initial stage of the operation, the surgeon manually introduces the bronchoscope into the trachea, and the remaining steps are completed by the robot.

The novel three-stage robotic bronchoscopy procedures described in chapter \ref*{procedure} is successfully performed and small tissue from the tertiary bronchus of the swine is sampled. The stiffness of the catheter is adjusted by energizing the resistance wire, providing adaptability based on the specific requirements of the procedure. The usability and effectiveness of the designed bronchoscopy surgical robot are demonstrated. During the biopsy sampling, the presence of the VS-Catheter makes the process more stable, and the implementation of the VS-Catheter enabled more flexible control of the biopsy forceps within the bronchus.

\section{Conclusion}
\label{Conclusion}

This paper introduces a teleoperated robotic bronchoscopy system and variable stiffness catheters to enhance the stability of robotic bronchoscopy for peripheral pulmonary lesion biopsy. The main innovations are the VS-Catheter integrated in robotic bronchoscopy system and the novel three-stage bronchoscopy procedure. The procedure includes initial insertion of bronchoscope robot, dynamic adjustment of VS-Catheter, and robotic biopsy. All the surgical processes are performed by remotely controlled robots.

The experimental study has validated the design concept and the feasibility of proposed robotic system. However, it is important to note that further validation is required through comprehensive preclinical studies and additional in-vivo tests involving surgeons. These subsequent evaluations will provide more insights into the practicality and efficacy of the system in real-life surgical scenarios.

\section*{Acknowledgments}
The study was approved by Institutional Review Board and Ethics Committee of Shenzhen Institutes of Advanced Technology (AAS 201205P). We thank Jie Jiang for helping in the preparation of animal experimental platform, Heyun Feng and Yaruo Zeng for the use of EM tracking system. 

\bibliographystyle{IEEEtran}
\bibliography{ref}

\vspace{12pt}
\color{red}

\end{document}